# Electric dipole moments of nanosolvated acid molecules in water clusters

Nicholas Guggemos[1], Petr Slavíček[2], and Vitaly V. Kresin[1]

[1]*Department of Physics and Astronomy, University of Southern California, Los Angeles, CA 90089-0484, USA*

[2] *Department of Physical Chemistry, University of Chemistry and Technology, Prague, Technická 5, 16628 Prague, Czech Republic.*

The electric dipole moments of $(H_2O)_n DCl$ ($n$=3-9) clusters have been measured by the beam deflection method. Reflecting the (dynamical) charge distribution within the system, the dipole moment contributes information about the microscopic structure of nanoscale solvation. The addition of a DCl molecule to a water cluster results in a strongly enhanced susceptibility. There is evidence for a noticeable rise in the dipole moment occurring at $n \approx 5\text{-}6$. This size is consistent with predictions for the onset of ionic dissociation. Additionally, a molecular dynamics model suggests that even with a nominally bound impurity an enhanced dipole moment can arise due to the thermal and zero point motion of the proton and the water molecules. The experimental measurements and the calculations draw attention to the importance of fluctuations in defining the polarity of water-based nanoclusters, and generally to the essential role played by motional effects in determining the response of fluxional nanoscale systems under realistic conditions.

*INTRODUCTION.* Water clusters are convenient experimental platforms for the study of the microscopic physics and chemistry of solvation. By monitoring cluster properties as a function of the number of water molecules, it is possible to follow the step-by-step progression of intermolecular interactions [1], following the transition from isolated molecule to bulk By virtue of having a large fraction of their molecules located near the surface, clusters can serve as surrogates for important processes occurring on aerosols and hydrometeors [2,3]. Composed of only a finite number of constituents, they often serve as test beds for theoretical methods and models.

Small water clusters doped with an acid molecule – here, for concreteness, hydrogen chloride – have been the subject of a great number of theoretical papers and a sizeable (but regrettably much smaller) number of experimental ones. Hydrogen chloride readily dissociates into $H_3O^+$ and $Cl^-$ and the dissociation also takes place in HCl hydrates [4-8], on the ice surface [9-12] or on larger water nanoparticles [13-15] and protonated water cluster ions [16,17]. One inquiry that persistently threads its way through this subject is: what is the minimum quantity of water molecules, $(H_2O)_n$, required to dissociate the acid [18]? According to calculations, the acidically dissociated structure is supported starting from $n=4$; most of the theoretical approaches agree that at this size the dissociated form represents the global minimum on the potential energy surface [18-24]. Experimentally, for neutral clusters the question has been probed by laser spectroscopy (see, e.g., [25-28]) but finding a conclusive spectroscopic signature of the onset of dissociation in a nanocluster is not straightforward.

As a matter of fact, the challenge is not just experimental but conceptual. The free water clusters in natural and laboratory environments generally exist at temperatures appreciably above absolute zero [29]. (Our focus is on this regime, as opposed to the environment in superfluid helium droplets [26-28].) To build a constructive bridge between theory and experiment, it is important to address the question: To what extent do parameters observable at finite temperatures either retain or lose the specificity assigned by optimization of the cluster structure? (See, e.g., Ref. [31] as well as [32-35].) In the bulk, dissociation of an acid molecule clearly implies separation of its charged constituents. But within a small water cluster the anion and the proton are confined to a finite volume, while the proton is highly mobile. Hence in the presence of thermal as well as zero-point motion it is not obvious to what degree ionic dissociation in a realistic cluster environment translates into an unambiguous change in the distribution of charge within the cluster. There exists a convenient observable which directly reflects electrical charge distribution within a system: the electric dipole moment ***p***. Indeed, measurements of |***p***| by electrostatic deflection of cluster beams have served as a valuable probe of structure and bonding [36-38]. However, there have been only few applications of the method to water clusters [39-41]. In the present report we use beam deflection to provide a new experimental angle onto the problem of acidic dissociation: a measurement of the electric dipole moment of $(H_2O)_n$ clusters carrying a DCl molecule. The measurements yield what may be the first direct evidence for a transition occurring between $n=5$ and 6.

The data are considered within the context of a theoretical analysis of proton delocalization and the associated time-averaged dipole moment (or, equivalently, the rovibrational polarizability) of a doped cluster. We point out that the dynamically averaged dipole moment within a finite highly fluxional system can be qualitatively different from that computed for a static minimum-energy framework.

*EXPERIMENTAL RESULTS.* Dipole moments were deduced from a beam-deflection measurements of the susceptibility (effective polarizability $\alpha_{\text{eff}}$) of clusters. Experimental details are described, in the Supplemental Materials section [42]. Mass spectrometric detection of water clusters doped with a molecule of hydrochloric acid is impeded by the fact that electron-impact ionization of the complex always results in a complete loss of the chlorine atom [54] making the mass spectrum indistinguishable from that of neat water clusters. To identify mass peaks deriving from doped species, it is therefore necessary to deuterate one of the partners. We used a supersonic oven filled with $H_2O$ vapor and injected DCl gas from a capillary into the expansion zone.

In a system whose dipole moment undergoes fluctuations and is statistically oriented along the electric field (often referred to as a "floppy" cluster), the probability of sampling a particular dipole configuration approximates a canonical distribution, and the effective linear susceptibility is then given by the Langevin-Debye formula with the temperature corresponding to the cluster's internal rovibrational temperature [36-38]: $\alpha_{\text{eff}} = \bar{p}^2 / (3 k_B T)$. Here $\bar{p} = |\bar{\boldsymbol{p}}|$ is the time-averaged value of the cluster dipole moment along the deflection field direction. It is important to emphasize (see, e.g., the incisive discussion in the classic book of Van Vleck [55]) that the numerator represents the statistical mean square of the vector dipole moment of the system (in the absence of an external field). It is a more general quantity than a permanent moment, and can be temperature-dependent. Thus it also incorporates the effects of rovibrational polarization and screening (cf., e.g., Ref. [56]). This correlates with the assertion above that discussions of the structure of molecular clusters must be supplemented by considerations of the dynamics of their constituents.

The experimentally measured effective polarizabilities are plotted in Fig. 1(a). The first important observation is that peaks assigned to mixed clusters display a much larger response, attesting to the presence of a polar impurity.

Water clusters formed in a hot nozzle expansion will cool by evaporation, and evaporative ensemble theory [57-59] predicts a resultant $T \approx 200$ K. This temperature was employed previously with the same setup to analyze the deflection of neat water clusters [39] and is close to $T_{\text{rot}}=167$ K fitted to the deflection profiles of the $D_2O$ molecule [60]. Using this value and the measured $\alpha_{\text{eff}}$, we obtain the root-mean-square (rms) values of the dipole moment $\bar{p}$ plotted in Fig. 1(b). The neat cluster dipole moments are approximately 10-20% higher than the values reported in an earlier measurement [39].

Fig. 1(b) shows that the excess dipole moment carried by the mixed clusters (i.e., the difference between the moment of a mixed and neat cluster) is close to that of the DCl molecule (1.1 D [61,62]). This implies that the peaks detected in the mass spectra do not originate from much heavier precursors: otherwise, the latter would have to possess unrealistically high dipole moments in order to deflect by the observed amount. This is consistent with the assumption that $(H_2O)_n$DCl does not undergo extensive fragmentation upon electron-impact ionization. Indeed, according to theoretical results [63] ionization of $(H_2O)_4$HCl dominantly involves the loss of zero or one water molecule; for larger clusters losses may be weaker. Similarly, Refs. [64,65] concluded that the number of water monomers lost by small neat water clusters upon electron bombardment or VUV radiation, respectively, is small. Thus we estimate that the uncertainty in the experimental dependence of cluster properties on size $n$ does not on average exceed one molecule. (The influence of fragmentation merits further investigation but calls for an ionization technique that would not perturb cluster structure and combine softness with efficiency.)

The data reveal a sizeable ≈20% increase in the electric dipole moment of $(H_2O)_n$DCl clusters between $n$=5 and $n$=6. In other words, at this point there occurs a restructuring of the average charge distribution within the cluster (since this is precisely what defines the dipole moment's magnitude). As remarked in the introduction, it is convenient that a deflection experiment can make such an effect directly apparent. This is the first direct observation of a shift in the electric susceptibility of a small doped water cluster with size.

CALCULATION AND DISCUSSION. A strong shift in the magnitude of $\bar{p}$ in a doped water cluster can be due to one or both of the following mechanisms: either a significant increase in the separation between $D^+$ and $Cl^-$ (i.e., dissociation of the molecule), and/or a strong shift in the rovibrational electric polarizability of the water cluster itself (i.e., a change in how effectively the water network screens the impurity). As emphasized in the introduction, a qualitatively important question in this context is the influence of internal motion on the actual dipole moment of a highly fluxional finite molecular cluster.

The important role played by quantum delocalization in the structure of $(H_2O)_n$HCl complexes has been addressed a number of times [31,33,35]; cluster structure also exhibits significant variations with temperature. Here, we aim to assess how zero-point and thermally driven structural variations are reflected in the effective electric dipole moments. To our knowledge, this specific angle has not been addressed in previous calculations on doped water clusters and therefore we have supplemented the experiment by calculations which pay particular attention to the difference between minimum-energy structures and systems that exhibit actual motional dynamics [42]. As a first step, such calculations were performed for two selected $(H_2O)_4$HCl clusters representing the covalent and ionic minima, as well as for pure water clusters. As described below, the calculations suggest that in these conditions the two aforementioned mechanisms – ionic dissociation and polarization of the water cluster matrix - can enhance cluster dipole moments by similar amounts, and it is not straightforward to separate their contributions. Furthermore, it becomes apparent that a "charge-separated structure" no longer entails full charge separation when thermal and zero-point fluctuations are taken into account.

The dipole moments of minimum energy structures of small pure and doped water clusters are shown in Fig. 2(a). For several of them, the finite-temperature rms dipole moments are also shown. Already for neat water clusters there are significant deviations between the "minimum-energy" dipole moments and the dynamical rms values. This is most dramatically exemplified by the tetramer. $(H_2O)_4$ in its fixed minimum-energy configuration has a zero moment due to symmetry (and as a result is, e.g., invisible in rotational spectroscopy [66-68] The effective moment measured in the present experiment is, however, $\bar{p} \approx 1.9$ D, reflecting the cluster's internal motion. Molecular- (MD) and path integral molecular-dynamics (PIMD) simulations yield an rms value of 1.40 D (taking into account the thermal effects at 200 K) and 1.78 D (considering also the vibrational delocalization) [69].

The analysis of doped water clusters is much more complicated due to their enormous structural diversity. This diversity together with the lack of information on the relative population and intercoversion rates of different isomers in the beam make it impractical, and to a degree unjustified at this stage, to address computationally the experimentally detected change of the dipole moment between the doped pentamer and hexamer structures.

It is instructive, however, to consider again the situation for the doped tetramer, $(H_2O)_4HCl$ [Fig. 2(a)]. This is the first size supporting the existence of the $H_3O^+ \ldots Cl^-$ ion pair in the minimum-energy configurations. Here looking at minimum-energy structures would suggest that the dipole moments of the ion pair structures should be almost double that of the covalent form. However, the differences of the PIMD rms dipole moments are substantially smaller, which means that the dynamic charge distributions become much more similar. This occurs because of the thermal and quantum motion within the cluster.

This highlights the important point that whereas minimum-energy depictions portray distinct and static covalent and ionic structures, in reality the light hydrogen atom travels rather freely within the finite cluster, as illustrated in Fig. 2(b). This conclusion is in accord with previous studies even though the particular structural distribution can depend on the electronic structure method used in the simulations and cluster temperature. This figure shows the probability of finding an H atom near the Cl atom for structures starting out as covalent and ionic configurations. Inclusion of quantum delocalization effects in PIMD points towards partial dissociation of the covalent structure and (even more conspicuously) a partially covalent character acquired by the nominally ion-pair structure. The two corresponding pairs of points marked by dashed circles in Fig. 2(a) show behavior which fully correlates with this observation: the nominally covalent form of the tetramer displays an enhanced rms dipole moment at finite temperature (reflecting partial dissociation), while the corresponding dipole moment of a "solvent-separated-pair" starting structure actually decreases somewhat (reflecting a decrease in the time-average charge separation).

The figure makes it clear that experimentally measurable quantities must be discussed in terms of hydrogen densities rather than precisely defined hydrogen positions. The "fluxional perspective" on the $(H_2O)_n$HCl clusters is in accord with previous studies [31,33,35], although of course the particular distribution of geometries is sensitive to the electronic structure level employed and to the assumed cluster temperature.

SUMMARY AND CONCLUSIONS. We reported on a measurement of the electric dipole moment of water clusters doped with a molecule of acid. The dipole moment is an important observable because it is directly related to the charge distribution within the system. This is the first such measurement on the archetypal system of a water cluster containing a hydrogen-halide molecule. The addition of a DCl molecule to an $(H_2O)_n$ cluster results in an overall strong enhancement of the dipole moment, and furthermore there is evidence for a significant rise in the dipole moment of $(H_2O)_n$DCl occurring at $n \approx 5\text{-}6$.

This size is consistent with predictions for the onset of ionic dissociation. Alternatively, a molecular dynamics model suggests that the dynamical rovibrational polarization of the cluster's water molecules can give rise to a large dipole moment even when the impurity remains nominally bound. This calculation draws attention to the fact that care is needed in relating the charge distribution in a finite and fluxional cluster to the degree of ionization of an embedded impurity. This is especially important for protons which can undergo strong thermal and zero-point fluctuations. Overall, the role of fluctuations in defining the electric susceptibility of water clusters has not been fully and quantitatively addressed in the literature, and merits in-depth theoretical analysis. Conversely, deflection experiments on pure and doped neutral water clusters can provide important information about their internal motion and about the relative populations and interconversions of structural motifs, thus supplying useful benchmarks for theoretical models. An especially important technical development would be the ability to control and vary the internal temperature of such neutral free clusters.

Note also that notwithstanding the underlying dynamical details, the observed jump in the effective dipole moment of doped water clusters should have important consequences, for example, for long-range interactions involving these clusters, e.g., for the stability of dipole-bound anions [70-72], and for the rates of cluster reactions with electrons and ions.

ACKNOWLEDGMENTS. We would like to thank Prof. Kit Bowen for very useful advice on the formation of mixed cluster beams and for productive discussions. Support of this work by the U.S. National Science Foundation (N.G. and V.V.K., Grant No. PHY-1068292) and the Czech Science Foundation (P.S., Project no. 14-08937S) is appreciated.

**Figure captions**

**Fig. 1**. (a) Effective polarizabilities of neat (bottom) and DCl-doped (top) water clusters. Straight lines and bands show the mean value and its standard deviation for the groups $n$=3-5 and $n$=6-9. (b) Neat (bottom) and doped (top) time-averaged (rms) electric dipole moments $\bar{p}$ obtained from the data in (a) using the Langevin formula, as described in the text. Straight lines show the mean value for the groups $n$=3-5 and $n$=6-9. There is a jump in the DCl(H$_2$O)$_n$ cluster susceptibilities between $n$=5 and 6.

**Fig. 2.** (a) Calculated dipole moments for (H$_2$O)$_{1-6}$ (diamonds) and HCl(H$_2$O)$_{0-6}$ (circles) systems. Open symbols correspond to minimum energy structures optimized at the *ab initio* Møller–Plesset (MP2) level, and filled symbols to *rms* moments $\bar{p}$ from path integral molecular-dynamics values at the assumed cluster temperature of 200 K. In the labels, COV stands for a covalently bound HCl in the cluster, CIP stands for a contact-ion pair, and SSP for a solvent-separated ion pair. Dashed circles highlight two pairs of doped tetramer structures for which minimum-energy and finite-temperature dipole moment values can be compared. See the text and Supplementary Materials [42] for additional information. (b) Density distribution of the closest Cl…H distance for the covalent and solvent separated ion pair of the HCl(H$_2$O)$_4$ cluster. Black dotted and solid curves: classical and PIMD simulations, respectively, for the covalent structure. Red dotted and solid curves: classical and PIMD simulations, respectively, for the ion pair structure.

**Figures**

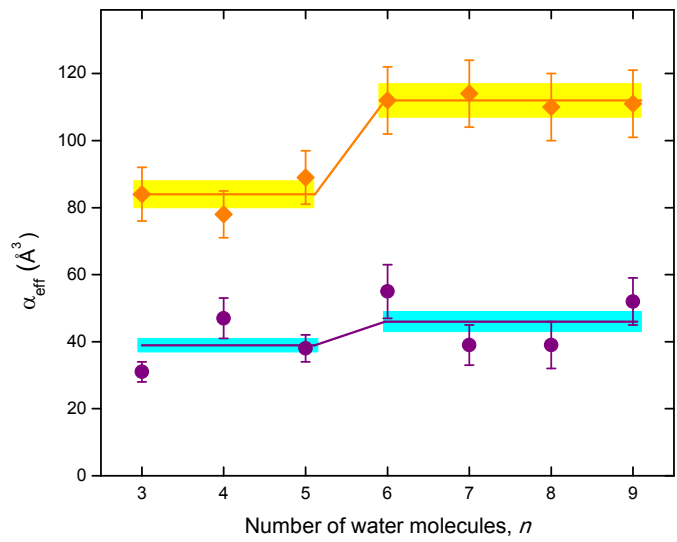

**(a)**

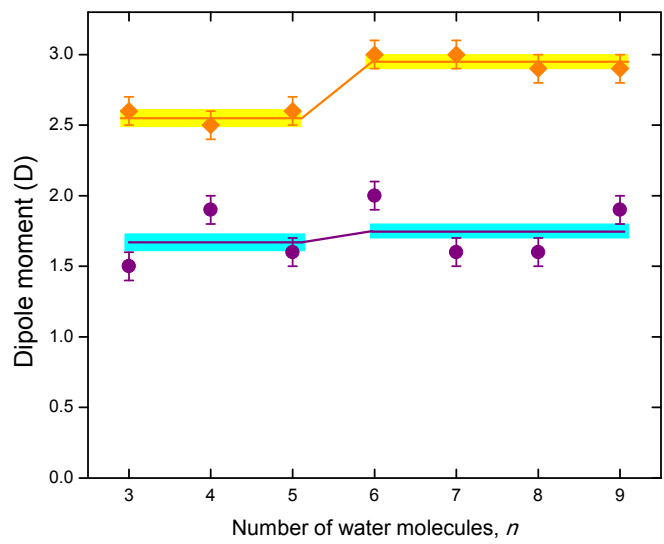

**(b)**

Fig. 1

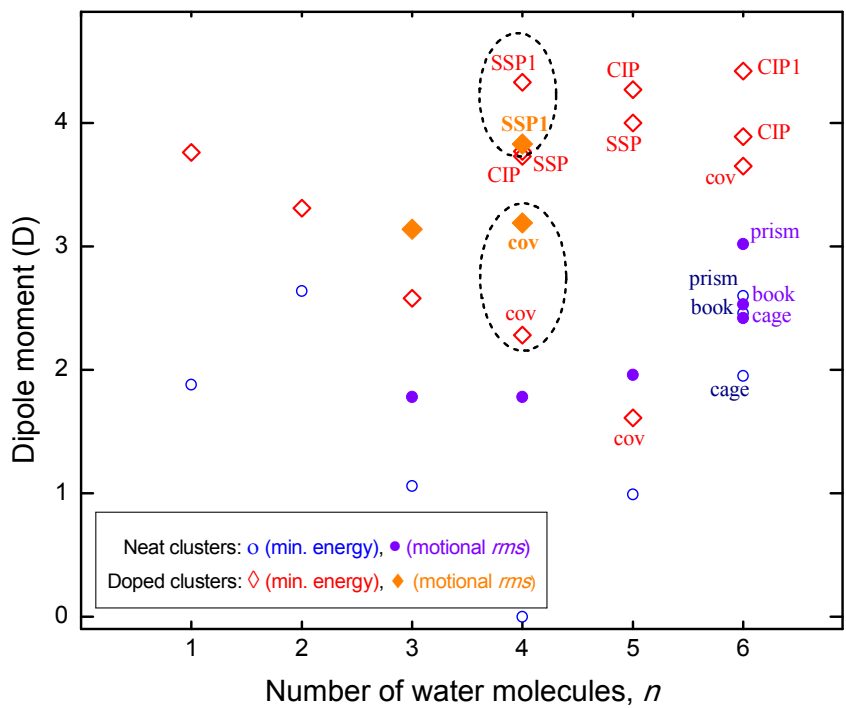

(a)

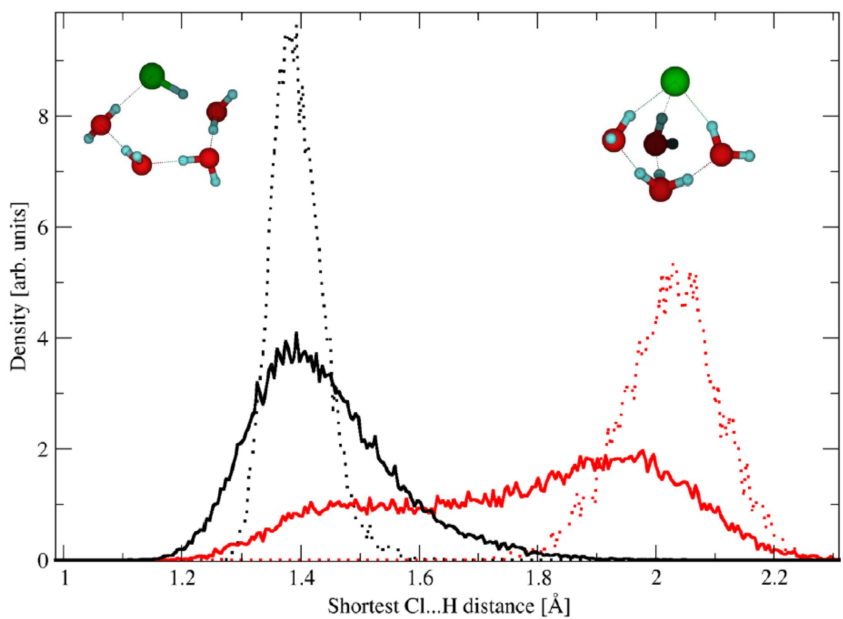

(b)

Fig. 2

SUPPLEMENTAL MATERIAL FOR

**Electric dipole moments of nanosolvated acid molecules in water clusters**

I. Experiment

The cluster beam apparatus was previously employed for electric deflection studies of neat water clusters [1] and molecules [2]. A diagram of the setup can be found in Ref. [1].

A beam of neutral $(H_2O)_n$ clusters is produced by heating distilled water in a stainless steel reservoir at 400 K and expanding its vapor through a 75 μm nozzle. As described in the main text, in order to identify mass peaks deriving from doped clusters it is necessary to deuterate one of the partners. Here we chose to use DCl and $H_2O$. In addition, it is necessary to introduce the acid molecules into the water cluster beam downstream from the source, because in a DCl/$H_2O$ solution D and H would become intermixed and substituted. A commonly used tool is a pick-up cell but we found that the resulting beam flux was much too weak after collimation for a deflection measurement. Instead, DCl gas (99%, Cambridge Isotope Laboratories) was injected directly into the expanding jet through a 1.6 mm capillary, slightly flattened at the tip, pointed at the nozzle from approximately 1 mm away. This produced a robust flux of singly-doped clusters.

Past a skimmer, the beam was collimated to 0.25×1 mm, and entered the gap between two 15 cm long metal plates shaped to create a "two-wire" electric field [3]. The field polarized the cluster and its gradient exerted a deflecting force on the oriented dipole. The beam was then chopped at 163 Hz by a rotating wheel before traveling for ≈70 cm to a quadrupole mass analyzer ("QMA": model UTI-100C, 0-300 amu range) with an electron-impact ionizer operated at 70 eV. The QMA had been upgraded to a pulse-counting channeltron multiplier (DeTech Inc.) and combined with a LabVIEW-based digital acquisition and control system [4,5]. A portion of the mass spectrum is shown in Fig. S.1. The most intense peaks came from protonated water clusters, because the main electron impact ionization channel is accompanied by the loss of a hydroxyl group [6,7]. The neighboring deuterated peaks derived from clusters on which a DCl molecule rode through the deflection path.

Deflections were measured by setting the QMA to the center of the corresponding mass peak. A 0.25 mm wide slit was scanned in front of the QMA ionizer entrance by a stepper motor. The beam profile was sampled at 15 slit positions in a sequence randomly selected for each pass across the beam. Background counts were subtracted by synchronizing the digital counters with the beam chopper. At each slit position, mass-selected signals were accumulated both with $V$=28 kV across the deflection plates and with the voltage off. Typically 6 to 12 passes were combined, after normalizing the intensity of each pass, into a final deflection profile, for a total acquisition time of 95 to 190 min. Examples are shown in Fig. S.2, together with a Gaussian fit.

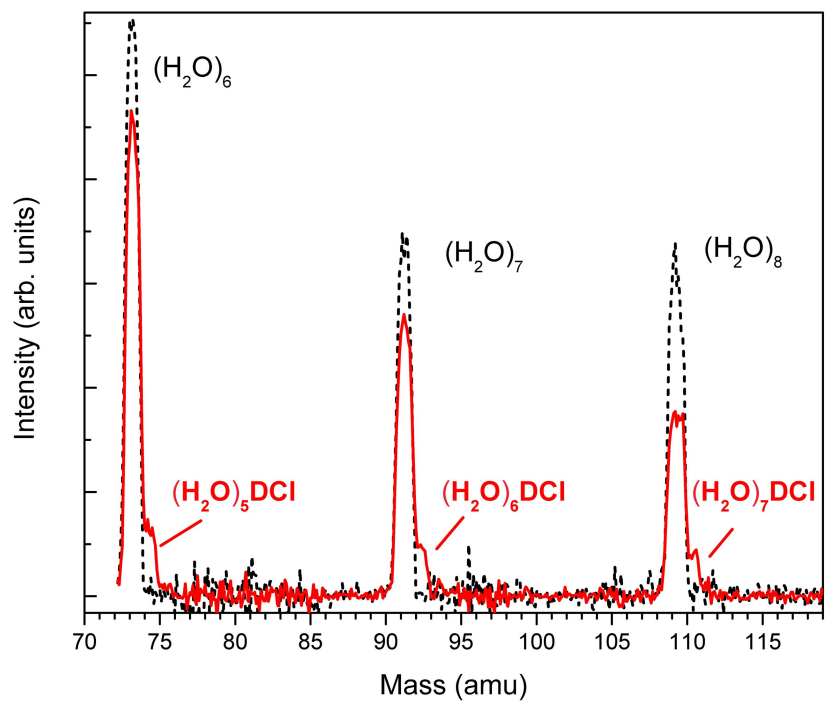

**Fig S.1** Segments of two mass spectra, with (solid) and without (dashed) DCl pickup by the water cluster beam. The peaks corresponding to doped cluster are well separated, permitting a measurement of their deflections.

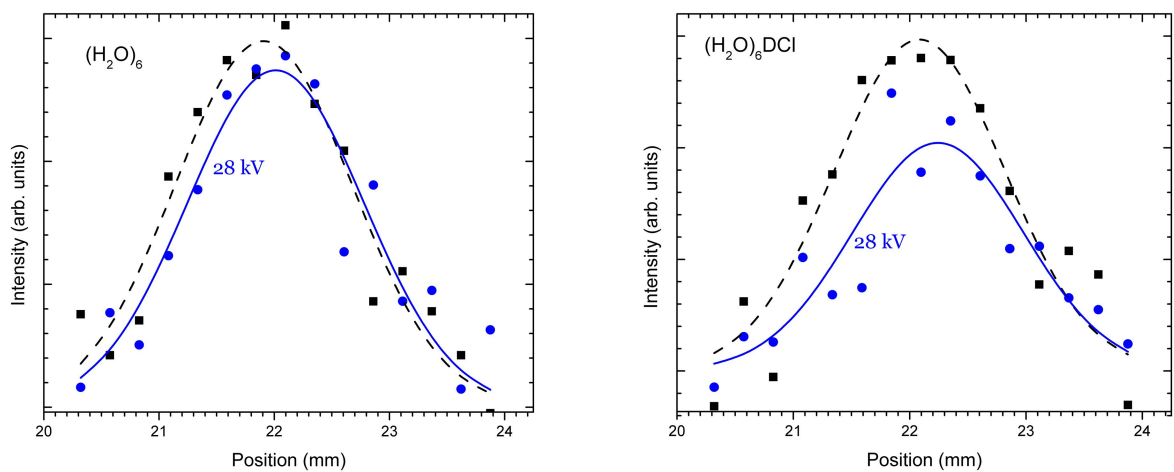

**Fig S.2** Sample deflection profiles of neat (left) and doped (right) clusters. Red: deflection field off, blue: deflection field on. The dots are beam intensities measured for each slit position, and the lines are Gaussian fits.

The amount of deflection, $d$, is proportional to $\bar{p}V/(mv^2)$ where $m$ is the cluster mass, $v$ its average velocity (from 900 to 960 m/s for doped clusters of size $n$=3-9) and $\bar{p}$ is the time-averaged value of the cluster dipole moment along the direction of the field gradient, the latter proportional to the applied voltage $V$. The above can be rewritten in terms of an effective polarizability, i.e., the ratio between the applied electric field (likewise proportional to $V$) and the cluster's dipole moment: $d = \alpha_{eff}CV^2/(mv^2)$. $C$, is the instrument's geometrical coefficient, was calibrated by the deflection of a beam of Kr atoms.

The measured $\alpha_{eff}$ for peaks assigned to $(H_2O)_n$ and $(H_2O)_n DCl$, are shown in Fig. 1(a) of the main text. The data for each cluster is the weighted average of deflections from 5-8 profile measurements as described above. The sum of the electronic polarizabilities of water clusters ($\approx 1.2n$ Å$^3$ [8]) and DCl (2.8 Å$^3$ [9]) is much less than $\alpha_{eff}$, hence the deflection is governed by the structural dipole moment of the cluster.

The error estimates used the method of statistical bootstrapping []. An experimental run consisting of $n$ beam profile scans produces a data file with $n$ cluster count values for each of the $m$ slit positions. Out of these, another group of $n$ is generated for each of the $m$ positions by picking the values at random with replacement, and the mean of this set is calculated. By carrying out this procedure for each slit position, one can build a full "synthetic beam profile." From every data file we constructed 300 such synthetic profiles and for each one calculated a Gaussian fit parameter (centroid, width and weight). The variances of these sets of 300 parameters were taken to represent the uncertainties in the actual deflection parameters for the run.

## II. Calculations

The results of the calculations are collected in Table S.1.

Dipole moments were first calculated for the minimum energy structures. The structures of $(H_2O)_{1-6}$ and $HCl(H_2O)_{1-6}$ were optimized at MP2/aug-cc-pVDZ level and the dipole moments were calculated at both the MP2/aug-cc-pVDZ and BLYP/6-31+g* levels. The former are in good agreement with the experimental values for the smallest systems. The latter deviate somewhat, but are included because the larger scale molecular dynamics calculations described in the next paragraph are only feasible with the DFT methods.

To account for the thermal and quantum fluctuations, we have calculated dipole moments of the respective clusters along Molecular Dynamics simulations. We have performed classical molecular dynamics simulations at 200 K (using Nosé-Hoover thermostat). The equations of motion were integrated with Velocity Verlet integrator using a timestep of 0.5 fs. The total duration of the simulations was 24 ps. To account also for the quantum effects, we have used the Path Integral Molecular Dynamics simulations. While such approach is guaranteed to converge to a correct quantum limit with increasing number of beads, the convergence can be rather slow. We took the advantage of the rapid acceleration of the convergence with respect to the number of beads using mode specific thermostatting within the framework of Generalized Langevin Equation (PI-GLE) [11]. This approach allowed us to perform the simulations with only 4 beads while still reaching the quantum distributions. The Reversible Reference System Propagator Algorithm (RESPA) was used to integrate the resulting equations of motion [12].

The dynamical simulations were performed on-the-fly on the DFT/BLYP/6-31+g* potential energy surface. The dipole moments were also recalculated at this level. All the electronic structure calculations were performed within Gaussian09 code [13].

**Table S.1.** Dipole moments for $(H_2O)_{1-6}$ and $HCl(H_2O)_{0-6}$ systems. The minimum energy structures were optimized at the *ab initio* MP2 level and the dipole moments calculated at the MP2 and DFT levels. The finite-temperature columns show molecular- and path integral molecular-dynamics values of $\bar{p}$ at the temperatures shown, scaled by a factor of 1.13 to account for the MP2/DFT difference in the minimal-energy structures. CIP stands for a contact-ion pair, SSP for a solvent-separated ion.

| | | Minimum energy structure dipole (D) | rms dipole (D), T=200 K | | | Minimum energy structure dipole (D) | rms dipole (D), T=200 K |
|---|---|---|---|---|---|---|---|
| | | | | HCl | | 1.47[a], 1.18[b]; *1.11*[c] | |
| $(H_2O)_1$ | | 2.23[a], 1.88[b]; *1.85*[c] | | $HCl(H_2O)_1$ | | 4.24[a], 3.76[b] | |
| $(H_2O)_2$ | | 2.92[a], 2.64[b]; *2.64*[c] | | $HCl(H_2O)_2$ | | 3.73[a], 3.31[b] | |
| $(H_2O)_3$ | | 1.21[a], 1.06[b] | 1.38[d], 1.78[e] | $HCl(H_2O)_3$ | | 2.83[a], 2.58[b] | 2.83[d], 3.14[e]; 2.65[d] (100 K) |
| $(H_2O)_4$ | | 0.00[a,b] | 1.40[d], 1.78[e] | $HCl(H_2O)_4$ | CIP | 3.44[a], 3.73[b] | |
| | | | | | SSP | 3.29[a], 3.77[b] | |
| | | | | | SSP1 | 4.07[a], 4.33[b] | 4.53[d], 3.83[e] |
| | | | | | covalent | 2.47[a], 2.28[b] | 3.03[d], 3.19[e] |
| $(H_2O)_5$ | | 1.12[a], 0.99[b] | 1.51[d], 1.96[e] | $HCl(H_2O)_5$ | CIP | 4.10[a], 4.27[b] | |
| | | | | | SSP | 3.56[a], 4.00[b] | |
| | | | | | covalent | 1.72[a], 1.61[b] | |
| $(H_2O)_6$ | Cage | 2.17[a], 1.95[b] | | $HCl(H_2O)_6$ | CIP | 3.61[a], 3.89[b] | |
| | Prism | 2.90[a], 2.60[b] | | | CIP1 | 4.53[a], 4.42[b] | |
| | Book | 2.73[a], 2.46[b] | | | covalent | 4.09[a], 3.65[b] | |

(a) DFT BLYP/6-31+g*;  (b) MP2/aug-cc-pVDZ;  (c) Experiment [9,14];  (d) MD;  (e) PIMD